# Measurement of Fluctuation-induced Diamagnetism in BSCCO-2212 single crystals using Magneto-optics

Huseyin Sozeri[*], Lev Dorosinskii and Ugur Topal

*TUBITAK-UME, National Metrology Institute, PO Box 54, TR-41470 Gebze-Kocaeli, Turkey*

Fluctuation induced diamagnetism in BSCCO single crystals with different doping levels ( i.e., different oxygen stoichiometry ) was measured using magneto-optics (MO) in a wide frequency range up to several MHz. A shift in $T_c$ onset, measured by MO, of up to 6 K towards higher temperatures was observed at high frequencies in those samples which are far from the optimally doped condition. An explanation of the observed effect in terms of the phase fluctuations of the superconducting order parameter is proposed.

[*]Corresponding Author: Tel:+90-262-679 5000, Fax:+90-262-679 5001

E-mail address: huseyin.sozeri@ume.tubitak.gov.tr (H. Sozeri)

**Introduction**

One of the important differences between HTSC and conventional superconductors is the strength of fluctuations of the order parameter. Both high anisotropy and high $T_c$ combined with the very short coherence length make the effect of fluctuations in cuprate superconductors very significant. The strength of fluctuations can be characterized by the Ginzburg number, $G_i$, which is higher in HTSC ($\approx 10^{-3} - 10^{-2}$) compared to that of conventional superconductors ($\approx 10^{-7} - 10^{-5}$). This causes the fluctuation controlled region to be several orders of magnitude wider in HTSC materials. Dimensionality of the system has also a very important effect, fluctuations are much stronger in low dimensional systems. As an example of the effect of the dimensionality, we can mention the so-called crossing point of magnetization in Bi-2212 [1-4], which is attributed to the strong fluctuations. Due to the different physical parameters of HTSC's as mentioned above, there is a wide region in between the normal and the superconducting state which is controlled by fluctuations of the order parameter. In addition to their unique superconducting parameters, HTSC's have a rich temperature-doping (T-δ) phase diagram [5] which affects both the superconducting properties and the strength of fluctuations.

The occurrence of fluctuations can be demonstrated experimentally by several ways. Creation of Cooper pairs with the characteristic life time of $\tau_{GL} = \dfrac{\hbar \pi}{8 k_B (T - T_C)}$ (in the vicinity of the transition) increases the electrical conductivity of the material above $T_c$ ( paraconductivity ). Although these pairs have a finite life time, a non-zero number of such pairs is always present in the normal state. Their presence not only gives rise to the appearance of the Meissner-Ochsenfeld diamagnetism in the normal state but also contributes to the heat capacity and other properties of the material. All these effects were successfully explained by the Aslamazov-Larkin (AL) theory of fluctuations [6] and it's extensions. However, recently new observations were made which can not be explained by this theory. Thus, in our earlier work [7], we observed fluctuational diamagnetism in underdoped YBCO at temperatures as high as 10 K above $T_c$. Later, effects of superconducting pairing were found even at higher temperatures. For example, diamagnetic domains were directly observed at temperatures far above $T_c$ in underdoped $La_{2-x}Sr_xCuO_4$ thin films using scanning SQUID microscopy [8]. In addition to that, Nernst measurements in the same type of samples showed that the signal remains at 50-100 K above $T_c$, from which they concluded that vortex-like excitations exist in the normal state [9,10]. Results of these nice experiments can not be explained by the classical theory of fluctuations for two reasons. First, the characteristic life time of the observed fluctuations is much longer than $\tau_{GL}$. Second, the width of the region controlled by fluctuations is very large which contradicts to the predictions of the classical AL theory.

Existence of pairing, or diamagnetism, in the temperature range $T_c < T < T^{MF}$, ( where $T^{MF}$ is the mean field critical temperature of the BCS theory and $T_c$ is the temperature where the phase coherence and thus the superconductivity, sets in), is well explained by the theory of Emery and Kivelson [11]. According to this theory, Cooper pairs are created below $T^{MF}$ and, although there is no phase coherence between them, they suppress the intensity of low energy excitations which looks like a pseudogap, which was directly observed in various experiments including NMR [12], neutron scattering [13], infrared conductivity [14], angle resolved photo emission [15] and scanning tunneling spectroscopy [16]. At a lower temperature, $T_c$, the macroscopic phase coherence sets in and material becomes superconducting.

In our study, magneto-optics (MO) is used to investigate the effect of the oxygen doping level on local ac magnetization measurements of Bi-2212 single crystals in a wide frequency range, between 5 Hz – 3MHz. Compared to traditional magnetization measurements, MO has certain advantages ( see our earlier works [7,17] and some review articles [18] ). Among those, adjustable measurement time scales may be the most important one for our purpose. Thus, MO enables us to do time-resolved studies of the dynamic properties of the superconducting state.

**Experimental**

BSCCO samples were grown by directional solidification method as explained in Ref. [19]. High purity $Bi_2O_3$, $SrCO_3$, $CaCO_3$ and $CuO$ (all %99 pure) were used as starting materials. Excessive $Bi_2O_3$ was included to act as a flux for the crystal growth. Samples with thickness of around 50 μm and having lateral dimensions of 0.6-0.8 mm in width and 1.5-2 mm in length were subjected to various heat treatments to obtain different oxygen stoichiometries. Annealing at 800 °C in air for 72 hrs followed by liquid nitrogen quenching, for example, gives nearly optimally doped BSCCO with $T_c$ of 92.3 K [20]. 10 days annealing at 400 °C with liquid nitrogen quenching results in underdoped samples with $T_c$ = 86.7 K. Finally, an overdoped sample having $T_c$ = 83.5 K was obtained by one-week annealing at 450 °C in a reduced oxygen atmosphere with the oxygen partial pressure of 0.0022 atm [21]. Transition temperatures of our crystals having different oxygen stoichiometry are close to the values published in Ref. [20,21] where similar heat treatments were done. $T_c$ values of our samples, were determined using MPMS-XL SQUID magnetometer by Quantum Design in the RSO mode and coincide with the values obtained in MO ( indicated as $T_c^{MO}$ ) at a low ac field frequency of 5 Hz.

Shielding of the ac magnetic field was measured using magneto-optic indicator films with in-plane anisotropy placed directly on top of the sample under investigation. Rotation of the magnetization vector in the indicator film causes deviation of the polarization vector in the incident polarized light. Thus, the field distribution, visualized due to the Faraday rotation of the light polarization, was observed in a polarizing optical microscope. Temperature dependences of the local ac susceptibility were obtained by measuring the oscillating light intensity with a photomultiplier connected to a lock-in amplifier [22]. Unlike ref. 22, where a special modulation technique used resulted in a signal directly proportional to the local DC field, AC field measurements done in the present work result in a signal approximately proportional to the Faraday rotation squared. As we are now interested in the transition onset only, we do not calculate the local AC susceptibility, but plot all curves in arbitrary units. The diameter of the area being sampled in our MO system is around 10 μm.

**Results & Discussion**

Temperature dependences of the local ac signal ( ~ local ac field amplitude ) of under-, optimally and overdoped BSCCO samples are shown in Fig.1a, b and c, respectively. Measurements were carried out in the frequency range from 5 Hz to 3 MHz with a fixed field amplitude of 5.5 Oe. The miniature ac coil is wound on a sapphire cold finger of an optical cryostat with a diameter of 9 mm. It's inductance is low enough, so that the AC field amplitude does not change with frequency up to 1 MHz. Since it is difficult to apply high amplitude of ac field at such high frequencies, our work is concentrated in the vicinity of $T_c$ ( at lower temperatures higher field amplitudes are required to achieve field penetration in the sample). It was observed that, for all doping levels, the transition widths are larger at low frequencies. For example, more than 15 K at 5 Hz for underdoped and around 9 K for overdoped BSCCO as seen in Fig. 1a and 1c respectively. As the frequency increases, the transition becomes narrower than 2 K for all doping levels. This broadening of the transition with decreasing the field frequency can be understood if flux creep is taken into account. According to the Bean model, magnetic induction decreases linearly from the sample edge, and the critical current corresponds to the slope of the B vs. x (distance from the edge) profile. Due to the creep, the actual slope is changed (i.e., decreased) according to the current relaxation during the period of the field oscillation. This decrease is stronger for lower frequencies, i.e., for longer creep times.

In addition to the transition widths, $T_c^{MO}$ onset values also change with the frequency, see Fig.2, which can not be explained by the creep. For the optimally doped sample, $T_c^{MO}$ onset is almost constant ( 92.7 K ) up to 100 kHz and then starts to increase at higher frequencies and reaches a maximum value of 95.4 K at 1 MHz. In under and overdoped (i.e. highly anisotropic) samples, on the other hand, the increase of $T_c^{MO}$ onset with frequency is more pronounced. There is a 6 K raise in the $T_c^{MO}$ onset of underdoped sample between 5 Hz to 3 MHz while it is 4.5 K and 3 K in the same frequency interval for the over- and optimally doped crystals respectively.

Before trying to understand this strange increase of the transition temperature, we performed some cross-checks to be sure that this is not an experimental error. First of all, the ac coil, which is wound directly on the cold finger, may change the temperature of the sample or the sensor. The heat released may result in overheating of the temperature sensor which will look like an increase in the transition temperature. However, the temperature of the sample space was measured with and without current in the ac coil and the difference appeared to be only 0.2 K at 5.5 Oe field amplitude, which is much smaller than the observed increase in $T_c^{MO}$ onset ( $\approx$ 3-6 K). Secondly, we made sure that all measurements were carried out at exactly the same location on the sample to exclude a possible effect of inhomogeneity. Moreover, we did measurements both on cooling and on heating the sample at several frequencies and no hysteresis was observed. Having understood that the observed shift in the transition temperatures is not an experimental artifact, we now come to a point to discuss the possible explanations of our results.

Our main observation is that the transition onset is above $T_c$ measured by the SQUID, and it shifts to even higher temperatures with increasing the frequency. This is a direct sign of the fluctuation induced conductivity. To make a quantitative comparison with the theory, we determined the frequency dependence of the conductivity by considering a simple skin model. Namely, what looks like a superconducting transition is in fact just shielding of the applied ac field. As the temperature decreases, the fluctuation-induced conductivity increases, that is why the shielding becomes stronger, i.e. the ac signal decreases. To calculate the AC conductivity we used the expression for a conducting sphere placed in an AC field [23]: $\alpha = \frac{1}{48\pi}\left[\frac{r}{\lambda}\right]^4$. Here $\lambda = \frac{c}{\sqrt{2\pi\omega\sigma}}$ is the skin depth, $c$ is the speed of light, $\omega$ is the field frequency, and $\sigma$ is the conductivity. The fact the our samples are not spherical will only result in a change of the numerical coefficient in the above expression. Then we can write down for the electrical conductivity: $\sigma = C\frac{\alpha^{1/2}}{A^2\omega}$, where $C$ is a constant and A is a characteristic size of the sample, approximately equal to its lateral dimension. Thus, we can find the conductivity $\sigma$ by measuring the frequency dependence of the susceptibility $\alpha$ at a fixed temperature. The dependence, shown in Fig. 3, is plotted in arbitrary units because the numerical coefficient in the above expression is not known (the above expression is correct as long as $\alpha \ll 1$, that is the penetration depth $\lambda$ is bigger than the sample size A, see [7] for a detailed discussion). These measurements were done at 3 different temperatures in the underdoped sample. As seen from the figure, the conductivity at a constant temperature changes with frequency even at low frequencies, which implies that our results can not be explained by the theory of AL [6] because, according to that theory, the conductivity should be independent of frequency up to $10^{11}$-$10^{12}$ Hz, the inverse of the lifetime of the AL fluctuations. Another contradiction with AL theory is the occurrence of strong fluctuations in a very wide temperature range, much wider than can be expected from the AL theory even for anisotropic superconductors. Both these observations, however, do not contradict to the theory of phase fluctuations by Emery and Kivelson. According to their theory, strong fluctuations of the phase of the superconducting order parameter destroy the phase coherence between Cooper pairs in the temperature range $T_c<T<T^{MF}$. The mean-field transition temperature, $T^{MF}$, can be quite high, especially in underdoped materials. Emery and Kivelson used this fact to explain the observations of the pseudogap at very high temperatures in underdoped samples. According to them, what people consider to be a pseudogap is in fact the usual superconducting gap but observed in the regime where the macroscopic superconducting coherence is destroyed by the phase fluctuations. A very similar situation was observed in the present work: the increase of the transition onset is the strongest in the underdoped samples. The fact that some weak increase was also observed for the optimally doped sample can mean that it is in fact not exactly optimally doped, but is close to that condition. The same conclusion is also supported by our earlier measurements on YBCO [7], where an increase of the transition onset with frequency (also very strong) was observed only in underdoped samples and by the SQUID microscopy [8] and Nernst effect [9,10] observations, which were also obtained on underdoped samples only. The fact that the transition onset, $T_c^{MO}$, shifts to higher temperatures with increasing the frequency is related with the shortening of the fluctuations lifetime at higher temperatures. That is, higher frequency measurements can detect fluctuations at higher temperatures. This would be not so in the case of the AL fluctuations because the characteristic frequency is too high, of the order of the inverse Ginzburg-Landau time, i.e. $\sim 10^{11}$-$10^{12}$ Hz. However, there is no yet theoretical derivation of the spectrum of the phase fluctuations. So, at least one cannot exclude the possibility that our low-frequency results are caused by these fluctuations. If it is really so, it is not surprising that the observed effect of frequency extends to very low frequencies. Finally, another fact which supports our conclusion is that the pseudogap was generally observed in underdoped [24-26] and, although weaker, in overdoped [16] samples, but never in optimally doped ones. That is both the pseudogap observations and our measurements are in fact observations of the phase fluctuations. In the theory developed by Emery and Kivelson, unfortunately there are no quantitative computations of the fluctuation spectrum in order to compare it with our experimental observations. Therefore, there is no threshold frequency up to which conductivity is frequency independent as in the case of AL theory.

To investigate effect of fluctuations at higher temperatures above $T_c$, we should apply higher frequencies, i.e. 100 MHz and even higher. To achieve such frequencies we are planning to do microwave absorption measurements on the same samples.

**Conclusion**

Superconducting fluctuations have been investigated using a new experimental technique, namely magneto-optics, in a wide frequency range in BSCCO-2212 single crystals having different doping levels. It was observed that the shielding of the ac field occurs in a wide temperature range above $T_c$ in under- and overdoped samples, while the transition is almost frequency independent for nearly optimally doped samples. These results support the recent model of phase fluctuations of the order parameter occurring below the mean field $T_c$ in under- and overdoped samples.

Finally, MO appears to be good method for detecting the effect of fluctuations in a wide frequency interval and studying their spectrum which is important for the comparison with existing theories.

**Figure Captions:**

Fig.1a, b and c: Frequency dependence of the local ac field for under-, optimally and overdoped samples respectively. Insets show fraction of the superconducting transitions for the first three lowest frequencies. $T_c$ onset values, shown by arrows, are used in Fig.2.

Fig. 2: Frequency dependence of the ac transition onset values for all doping levels. Error bars represent the uncertainty in $T_c$ onset determined as the temperature change which causes a change in the signal in graphs of Fig.1 equal to the noise value.

Fig.3: Frequency dependence of the conductivity for underdoped sample at different temperatures.

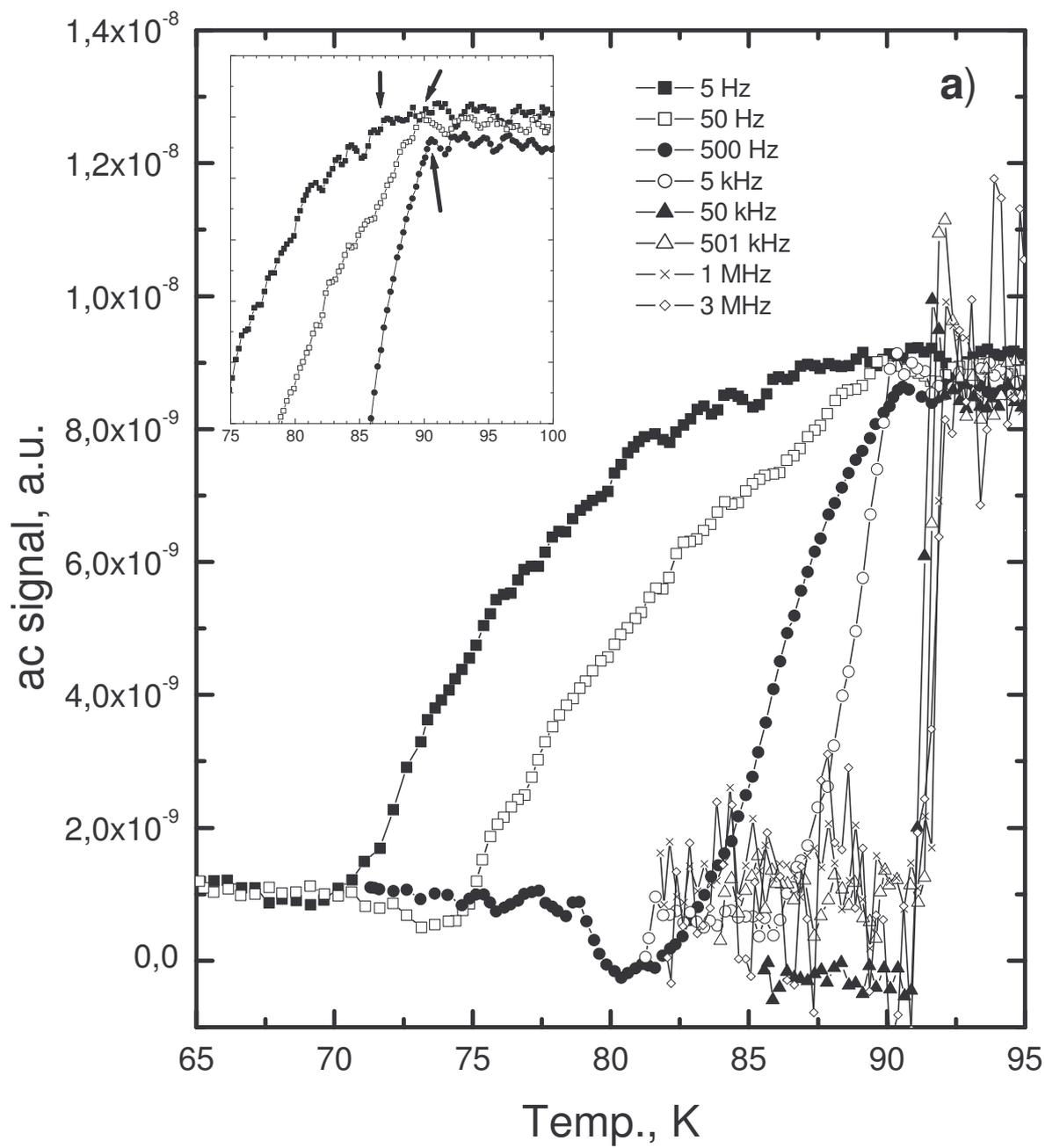

Fig.1

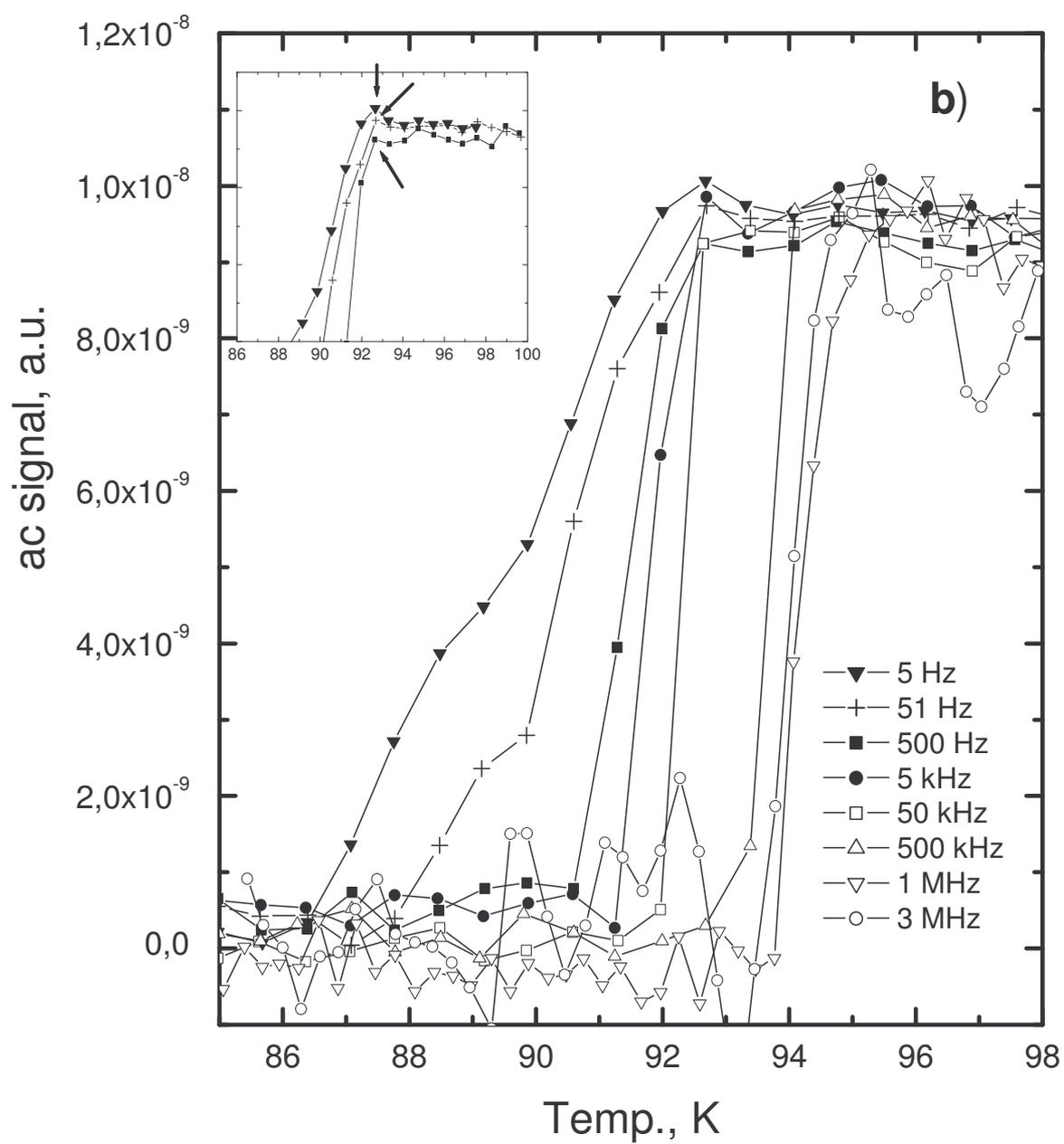

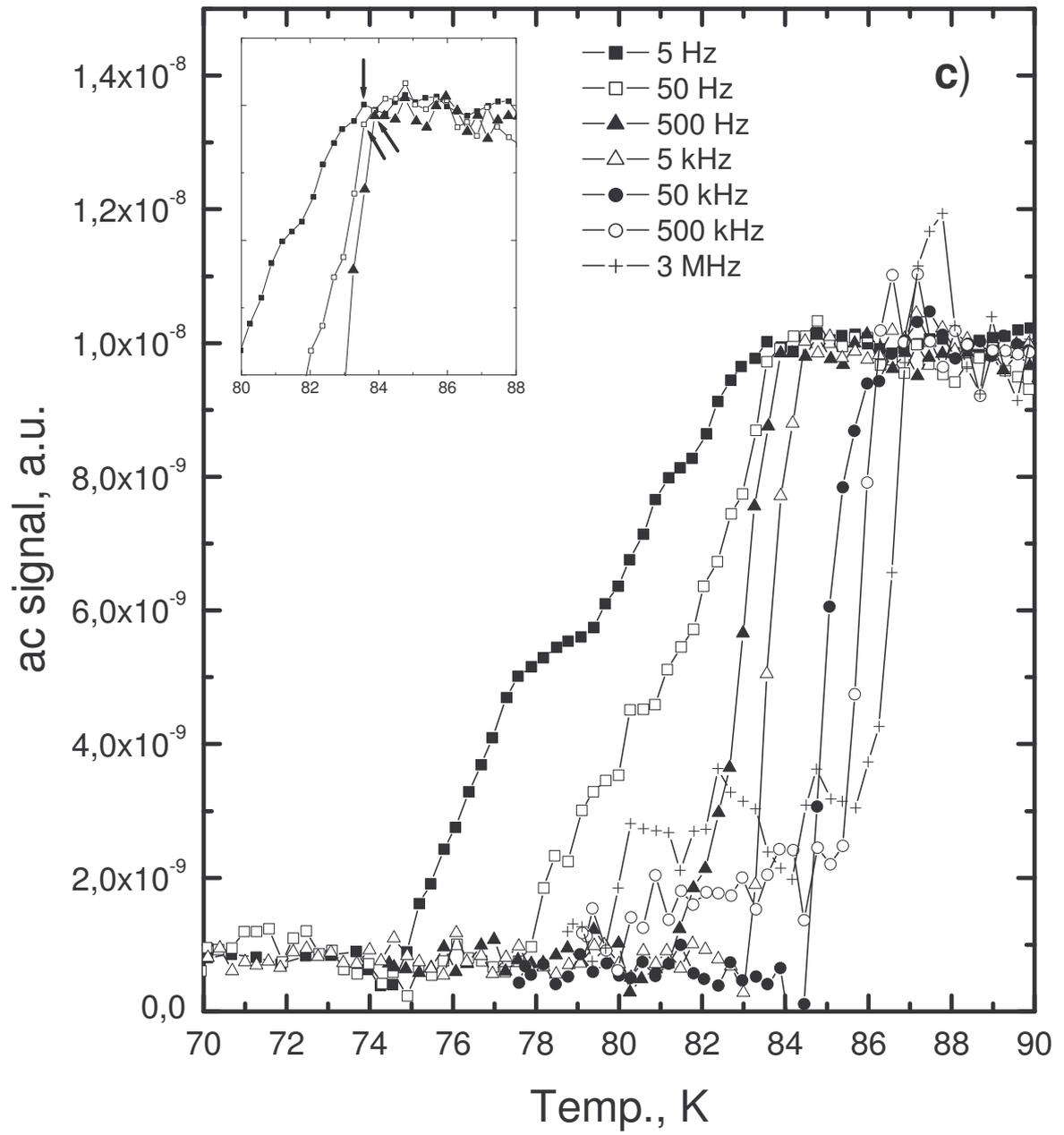

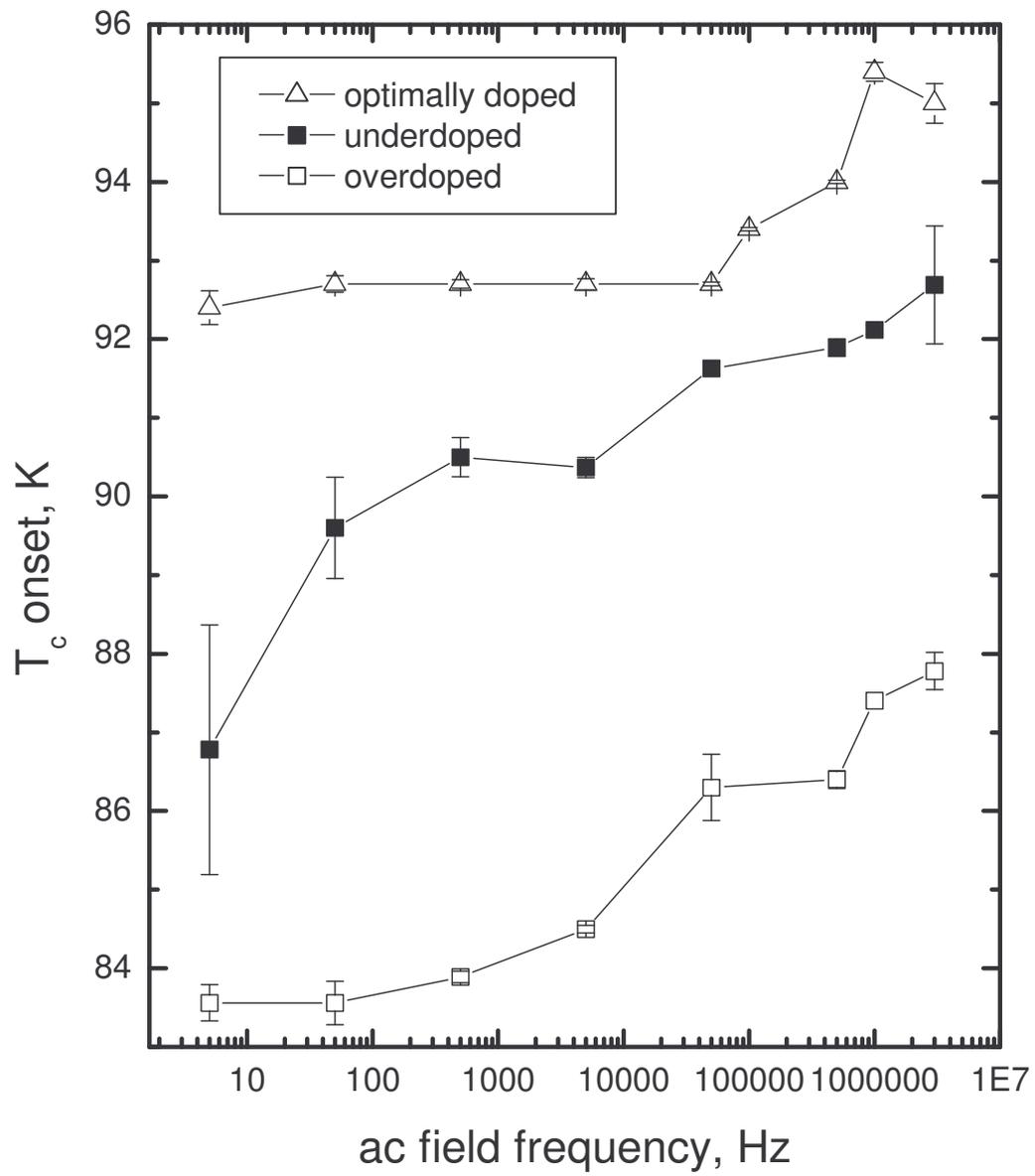

Fig.2

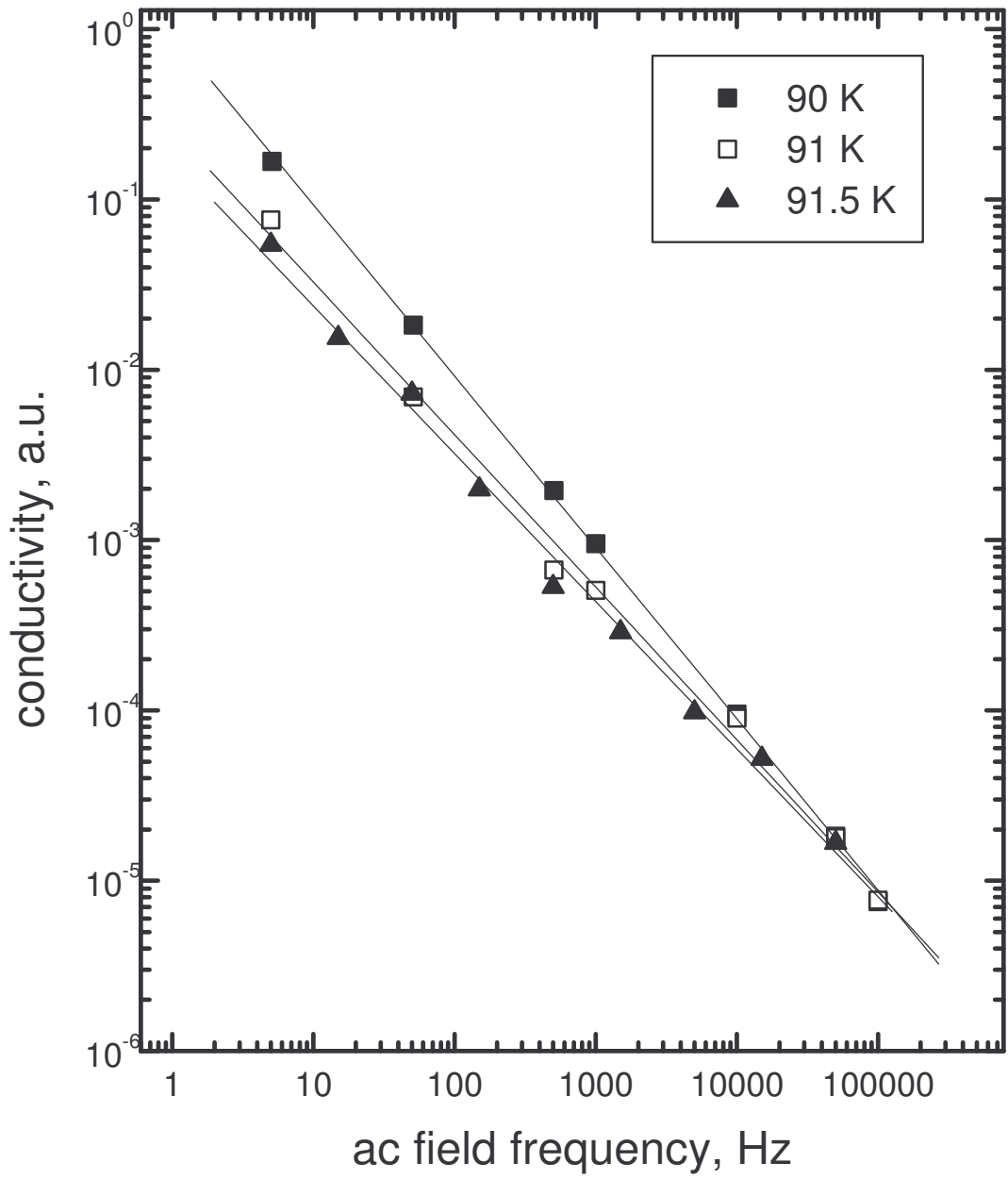

Fig.3